\documentclass[review,12pt]{elsarticle}

\biboptions{square,comma,sort&compress}

\usepackage{amssymb} 
\usepackage{amsthm} 
\usepackage{amsmath}
\usepackage{mathrsfs}
\usepackage{graphicx}
\usepackage{epstopdf}
\usepackage{float}
\usepackage{caption}
\usepackage{soul}
\usepackage{color,soul} 
\usepackage{color} 
\usepackage{subcaption}
\usepackage{bm}
\usepackage{bbm}
\usepackage{mathrsfs}
\usepackage{soul}
\usepackage[margin=2cm]{geometry}
\usepackage{booktabs}
\usepackage{chngpage}
\usepackage{enumitem}
\usepackage{subcaption}
\usepackage{multirow}
\usepackage{stackengine} 
\usepackage{mhchem} 
\usepackage{hyperref} 
\usepackage{cleveref}

\journal{International Journal of Hydrogen Energy}

\makeatletter
\def\@author#1{\g@addto@macro\elsauthors{\normalsize%
    \def\baselinestretch{1}%
    \upshape\authorsep#1\unskip\textsuperscript{%
      \ifx\@fnmark\@empty\else\unskip\sep\@fnmark\let\sep=,\fi
      \ifx\@corref\@empty\else\unskip\sep\@corref\let\sep=,\fi
      }%
    \def\authorsep{\unskip,\space}%
    \global\let\@fnmark\@empty
    \global\let\@corref\@empty  
    \global\let\sep\@empty}%
    \@eadauthor={#1}
}
\makeatother

\makeatletter
\def\thickhline{%
  \noalign{\ifnum0=`}\fi\hrule \@height \thickarrayrulewidth \futurelet
   \reserved@a\@xthickhline}
\def\@xthickhline{\ifx\reserved@a\thickhline
               \vskip\doublerulesep
               \vskip-\thickarrayrulewidth
             \fi
      \ifnum0=`{\fi}}
\makeatother

\newlength{\thickarrayrulewidth}
\begin{document}

\begin{frontmatter}



\title{On the relative efficacy of electropermeation and isothermal desorption approaches for measuring hydrogen diffusivity}


\author{Alfredo Zafra\fnref{IC}}

\author{Zachary Harris\fnref{Pitt}}

\author{Evzen Korec\fnref{IC}}

\author{Emilio Mart\'{\i}nez-Pa\~neda\corref{cor1}\fnref{IC}}
\ead{e.martinez-paneda@imperial.ac.uk}

\address[IC]{Department of Civil and Environmental Engineering, Imperial College London, London SW7 2AZ, UK}

\address[Pitt]{Department of Mechanical Engineering and Materials Science, University of Pittsburgh, Pittsburgh, PA 15261, USA}

\cortext[cor1]{Corresponding author.}

\begin{abstract}
The relative efficacy of electrochemical permeation (EP) and isothermal desorption spectroscopy (ITDS) methods for determining the hydrogen diffusivity is investigated using cold-rolled pure iron. The diffusivities determined from 13 first transient and 8 second transient EP experiments, evaluated using the conventional lag and breakthrough time methods, are compared to the results of 10 ITDS experiments. Results demonstrate that the average diffusivity is similar between the second EP transient and ITDS, which are distinctly increased relative to the first EP transient. However, the coefficient of variation for the ITDS experiments is reduced by 2 and 3-fold relative to the first and second EP transients, confirming the improved repeatability of ITDS diffusivity measurements. The source of the increased error in EP measurements is systematically evaluated, revealing an important influence of assumed electrochemical boundary conditions on the analysis and interpretation of EP experiments.
\end{abstract}

\begin{keyword}

Hydrogen diffusion \sep Electro-permeation \sep TDS \sep Isothermal desorption \sep repeatability  



\end{keyword}

\end{frontmatter}

\section{Introduction}
\label{Introduction}

The deployment and safe operation of hydrogen transport and storage infrastructure is being threatened by the degradation of metals when exposed to hydrogen \cite{Bouledroua2020,Cerniauskas2020,Eames2022}. However, despite over a century of study \cite{Johnson1875,Gangloff2012,Robertson2015,Djukic2019}, hydrogen-assisted cracking (HAC) remains a relevant failure mode for metallic structural components exposed to hydrogen-producing/containing environments across a number of industrial sectors \cite{Gangloff2003}. Current best practices for designing against and prognosis of HAC involve the use of linear elastic fracture mechanics (LEFM)-based damage tolerant design \cite{Gangloff2016,EFM2017,Valverde-Gonzalez2022}. While fundamentally robust, the implementation of LEFM-based approaches can be complicated by the sensitivity of HAC behavior to changes in microstructural \cite{Pallaspuro2017,Harris2019,Lin2022,Shoemaker2022}, mechanical \cite{Wang2005,Alvarez2019,Harris2021}, and environmental \cite{McMahon2019,Steiner2021,Gangloff2014} parameters. As such, it is critical that the influence of these factors be well understood when evaluating new alloy systems for use in hydrogen-rich environments.

From an environmental perspective, the parameters most pertinent to HAC are the diffusible hydrogen concentration ($C_{H,Diff}$) and the hydrogen diffusivity ($D$) in the material of interest \cite{SanMarchi2012,Oudriss2014,Zafra2018,IJHE2016}. Regarding the former, $C_{H,Diff}$ represents the hydrogen content available to participate in the hydrogen-assisted fracture process and is strongly dependent on the operative environment conditions (\textit{i.e.}, solution composition/pH, electrochemical potential, hydrogen pressure/fugacity, etc. \cite{Turnbull1989,Zakroczymski1975,Gangloff2012,Liu2014,Kittel2016,McMahon2019,Livia2022}). It is well-established that susceptibility to HAC is dependent on $C_{H,Diff}$. For example, the threshold stress intensity associated with the onset of hydrogen-assisted cracking (K$_{TH}$) has been observed to decrease as $C_{H,Diff}$ increases \cite{LeeGangloff2007,LiScullyGangloff2004,Kehler2008,Harris2016}; this dependence is explicitly incorporated into proposed models for K$_{TH}$ \cite{Akhurst1981,Gerberich2012,Emilio2016}. Similarly, the Stage II hydrogen-assisted crack growth rate has been correlated with hydrogen diffusivity across a range of relevant alloy systems \cite{Gangloff2008Proc,Colombo2015a,Shinko2021}, particularly under severe hydogen-producing conditions \cite{Harris2021}. As with K$_{TH}$, the hydrogen diffusivity is a critical parameter in existing models for the Stage II crack growth rate \cite{Gerberich2012,Gangloff2014}.

These direct linkages between hydrogen-metal interaction parameters and HAC metrics underscores the importance of characterizing these factors when assessing the compatibility of a new alloy for use in aggressive environments \cite{Gangloff2016Poc,Depover2021}. While all hydrogen-metal interactions are critical to understand and contribute to HAC susceptibility, the hydrogen diffusivity is of particular importance given that it (1) affects the rate at which a given concentration is obtained in the material, and (2) is explicitly required for the interpretation of some hydrogen-metal interaction experiments (\textit{e.g.}, barnacle cell electrode \cite{DeLuccia1981}). There are two primary experimental methods for determining the hydrogen diffusivity \cite{Depover2021}: permeation and thermal desorption. Permeation experiments are performed using a thin membrane of the material of interest that separates two environments. Hydrogen is generated on one side of the membrane, diffuses through the material, and exits the membrane at the other side, where the rate of hydrogen egress is measured \cite{Depover2021}. For electrochemical permeation (EP) experiments, hydrogen uptake is driven by cathodically polarizing one side of the membrane and then measuring the current induced by hydrogen oxidation on the egress side \cite{Subramanyan1981}. Standard analysis methods are then used to determine the hydrogen diffusivity from the permeation data \cite{ISO17081:2014}. While permeation generally involves starting with nominally hydrogen-free specimens, diffusivity measurements using thermal desorption methods utilize samples that have ideally been hydrogen pre-charged with a spatially uniform hydrogen concentration. Samples are then placed into an ultra-high vacuum system, which is heated to a specific temperature (\textit{i.e.}, isothermal conditions) and the rate of hydrogen egress is monitored with a mass spectrometer \cite{Yamabe2021,Mine2009}. The diffusivity is then determined through fitting the hydrogen egress rate versus time profile using either numerical \cite{IJHE2020} or analytical methods \cite{Mine2009}.

While both approaches have been widely utilized in the open literature, each approach has been historically leveraged for a specific subset of conditions. Isothermal desorption spectroscopy (ITDS) is commonly employed for slow-diffusing materials and often involves large specimens to minimize the effects of hydrogen egress while the sample analysis chamber is brought to ultra-high vacuum and the selected isothermal condition is reached \cite{Mine2009,Yamabe2015,Yamabe2021,Ai2013,Hurley2015}. Conversely, EP is generally employed on thin (\textless 1 mm thick) membranes of fast-diffusing material \cite{Subramanyan1981,Boes1976}. While data obtained from each method are often compared via literature sources, direct assessments of the relative efficacy of the two techniques are minimal given their use under distinct conditions. However, recent advances in thermal desorption systems, such as the introduction of conduction-based heating and improved vacuum system design for more rapid sample introduction, now enable the use of ITDS for material/sample combinations that were previously considered incompatible with this technique. For example, Zafra \textit{et al.} \cite{Zafra2022} recently demonstrated that ITDS and EP methods yielded similar average diffusivities for thin (\textless 1 mm thick) sheet specimens of cold-rolled pure Fe. Critically, this direct comparison revealed that the error in  replicate ITDS measurements was noticeably reduced relative to the replicate EP experiments \cite{Zafra2022}. It is well-known that EP experiments are prone to significant scatter, which has been historically attributed to numerous factors: the need for sample conditioning prior to starting the permeation experiment \cite{Xie1982,Zakroczymski1985,Chen1992,Bruzzoni1999,Liu2014,Zhou2018}, surface effects \cite{Kiuchi1983,Addach2009}, trapping effects \cite{Kumnick1974,Lee1987,Pundt2006}, concentration-dependent diffusion \cite{Ono1968,Zafra2020}, analysis method assumptions \cite{Boes1976,Carvalho2017,Montella1999,Pound1993}, failure to reach true steady-state conditions \cite{Nelson1973}, and test-to-test variations in specimen thickness, environmental parameters, among other \cite{Gonzalez1969,Turnbull1995}. The work of Zafra \textit{et al.} \cite{Zafra2022} suggests that this longstanding issue of error in EP experiments may be circumvented via the more widespread adoption of ITDS, but this prior study only performed two experiments per technique. As such, additional experiments are needed to more rigorously compare the ITDS and EP methods for measuring hydrogen diffusivity.

The objective of this study is to provide the first statistically significant comparison of EP and ITDS approaches for measuring hydrogen diffusivity. A total of 21 permeation experiments are performed on cold-worked pure iron and then analyzed using the standard breakthrough and lag time methods \cite{ISO17081:2014}. The EP experiments are then compared to the results of ten ambient temperature ITDS experiments. These large datasets are subsequently leveraged to comment on the relative accuracy of each technique, the importance of assumed boundary conditions when analyzing EP data, and the broader implications of these findings for the hydrogen community. The results obtained reveal a higher sensitivity of ITDS measurements and demonstrate its applicability to fast diffusion materials.

\section{Experimental Methods}
\label{Sec:Experimental}

\subsection{Material}
This study was conducted using cold-rolled pure iron procured in the as-rolled condition from Goodfellow Ltd. as a 1-mm thick sheet. The supplier reported a purity of $>$99.5 wt. \%Fe and a 50\% reduction as the average degree of cold work. Permeation and TDS experiments were performed on plate specimens with nominal dimensions of 250 mm x 250 mm x 1 mm and 10 mm x 10 mm x 1 mm, respectively. The specimens were excised from the sheet using an abrasive saw. In order to ensure consistent surface conditions between experiments, both faces of every sample were mechanically ground using SiC papers, finishing with 1200 grit. 

\subsection{Hydrogen permeation tests}
\label{Subsec:HydrogenPermeation}
Thirteen EP experiments were performed at room temperature (22$\pm$1$^\circ$C) using a modified Devanathan-Stachurski double-cell. The specimen thickness, $L$, was 0.92$\pm$0.02 mm and a circular area of 2 cm$^2$ (16-mm diameter) was exposed to both sides of the double-cell.

The hydrogen reduction cell was filled with 3 wt. \% NaCl solution and contained a typical three-cell electrode system, with a Pt counter electrode, silver/silver chloride (Ag/AgCl) reference electrode, and the sample as the working electrode. Hydrogen production was achieved by applying a cathodic current density, $J_{c}$, of 5 mA/cm$^2$ to the cold-rolled Fe membrane using a Gamry 1010B potentiostat operated in galvanostatic mode. The corresponding cathodic potential (vs. Ag/AgCl) was measured at the beginning of the test to ensure consistent charging conditions between specimens. The hydrogen oxidation cell was filled with 0.1 M NaOH solution and also contained a Pt counter electrode and Ag/AgCl reference electrode, with a second Gamry 1010B potentiostat operated in chronoamperometry mode to record the permeation current density, $J_p$, as a function of time. Both solutions were actively deaerated via bubbling with pure nitrogen, which was initiated 1 hour before the beginning of the permeation test and continued through the end of the experiment. For each experiment, a thin Pd layer was electroplated onto the specimen surface facing the oxidation side of the double-cell to enhance the hydrogen oxidation reaction kinetics and avoid disturbances in the permeation signal from iron oxidation \cite{Devanathan1962,Driver1981,Manolatos1995}.

To ensure the efficient oxidation of hydrogen atoms reaching the anodic side of the specimen, the anodic surface was polarized at -25$\pm$11 mV$_{Ag/AgCl}$, which corresponds to the nominal open-circuit potential (OCP) for this environment. The permeation current density was then allowed to stabilize until reaching a baseline below 0.1-0.2 $\mu$A/cm$^2$, after which the galvanostatic cathodic charging was started on the reduction side of the double-cell. A hydrogen concentration gradient is thus generated in the specimen, with hydrogen atoms permeating through the iron membrane from the cathodic to the anodic side. During the test, $J_{p}$ describes an exponential rising transient until reaching a maximum permeation current density, which is commonly known as the steady-state permeation current, $J_{ss}$. Further details regarding the EP procedure as well as a schematic representation of the experimental setup (including the reduction and oxidation reactions) are provided elsewhere \cite{Zafra2022}. After completing the first permeation transient ($J_{c}$=5 mA/cm$^2$), a second transient was performed on eight of the samples by increasing $J_{c}$ to 10 mA/cm$^2$. These experiments were performed to assess whether the variability in diffusion coefficient would improve under conditions where surface and trapping effects on permeation are reduced \cite{Zafra2020a}. 

Consistent with precedent literature \cite{Boes1976,Turnbull1989,Luppo1991,Wu1992,Frappart2011,Svoboda2014,VandenEeckhout2017,Drexler2020,Zafra2020a,Zafra2020b} and current permeation testing standards \cite{ISO17081:2014,ASTMG148}, the hydrogen diffusion coefficient, $D$, of both permeation transients was determined using the breakthrough time and lag time methods, which are based on the following relationship:
\begin{equation}\label{eq:Diffusivity}
    {D}=\frac{L^2}{Mt}
    \end{equation}
    
\noindent where $t$ and M are determined by which method is being employed. For the breakthrough method, M has a value of 6 and $t$ is defined as the breakthrough time, $t_{bt}$, which corresponds to the time where $J_p$/$J_{ss}$=0.1. Conversely, for the lag time method, M is equal to 15.3 and $t$ is defined as the lag time, $t_{lag}$, which is determined by the time when $J_p$/$J_{ss}$=0.63. Both approaches represent closed-form solutions of Fick's second law under the assumption that that the hydrogen subsurface concentration is a constant, finite value at the entry side and zero at the exit side of the membrane.

\subsection{Isothermal TDS tests (ITDS)}
\label{Sec:IsothermalTDStests}

Ten ITDS experiments were performed on 0.90$\pm$0.02 mm thick specimens using a thermal desorption spectroscopy (TDS) system. Each specimen was first pre-charged with hydrogen in deaerated 3 wt. \% NaCl solution using a Gamry 1010B potentiostat operated in galvanostatic mode to maintain a constant cathodic current density of 5 mA/cm$^2$. The specimens were charged for 3 hours to obtain a nominally uniform hydrogen concentration across the plate thickness, based on the previously reported hydrogen diffusivity for the current material heat \cite{Zafra2022}. ITDS measurements were performed under ultra-high vacuum (10$^{-9}$ mbar) conditions using a previously described TDS system \cite{Zafra2022} equipped with a regularly calibrated Hiden Analytical RC PIC quadrupole mass spectrometer with a hydrogen detection resolution of $4.4\times10^{-6}$ wppm/s. The elapsed time between the completion of cathodic charging and the beginning of the ITDS experiment was 25 minutes. This is the minimum time needed to clean and dry the sample, load it into the system, and reach ultra-high vacuum conditions. All ITDS experiments were conducted at ambient temperature, which varied slightly across the test matrix (22.5$\pm$0.5$^\circ$C), though the measured variation in temperature during a given individual experiment was always below $\pm$0.1$^\circ$C. The rate of hydrogen egress (wppm/s) from the sample was then measured as a function of time for 6000 seconds.

The hydrogen diffusion coefficient, $D_{iso}$, was determined by fitting a 1-D finite element (FE) simulation of hydrogen transport to the measured ITDS profile (wppm/s vs. time) for each experiment. Recall that diffusion is governed by Fick's second law, which in a one-dimensional form reads:

\begin{equation}
    \frac{\partial C}{\partial t} = D \frac{\partial^2 C}{\partial x^2}
\end{equation}

\noindent where $C$ is the diffusible hydrogen concentration. For the current study, it is assumed that a homogeneous hydrogen concentration exists across the specimen thickness at the start of the simulation (\textit{i.e.}, $C=C_{0iso}$ at $t=0$) and that the hydrogen concentration at the free surface is zero given the use of an ultra-high vacuum environment (\textit{i.e.}, $C=0$ at $x=\pm L/2$). A FE mesh sensitivity study demonstrated that mesh-independent results were efficiently obtained for a progressive mesh composed of 200 elements decreasing in size towards the free surface of the specimen. As all experiments were performed at ambient temperature (22$^\circ$C) and did not use the heater stage of the TDS, $D_{iso}$ was determined in a single step analysis through fitting the experimentally-measured hydrogen desorption rate versus time data by iteratively changing the $C_{0iso}$ and $D_{iso}$ values. FE calculations were performed using the commercial package COMSOL Multiphysics and the LiveLink tool was employed to integrate COMSOL with Matlab, where a routine for parameter optimization was developed using the curve fit function \texttt{lsqcurvefit}, which performs a least square minimization using the Levenberg-Marquardt algorithm \footnote{The COMSOL model and the Matlab fitting subroutine are made freely available at \url{www.empaneda.com/codes}.}. 

\section{Results}
\label{Sec:Results}

\subsection{Permeation tests}
\label{Subsec:ResultsPerme}

\begin{figure}[H]
     \centering
         \centering
         \includegraphics[width=0.8\textwidth]{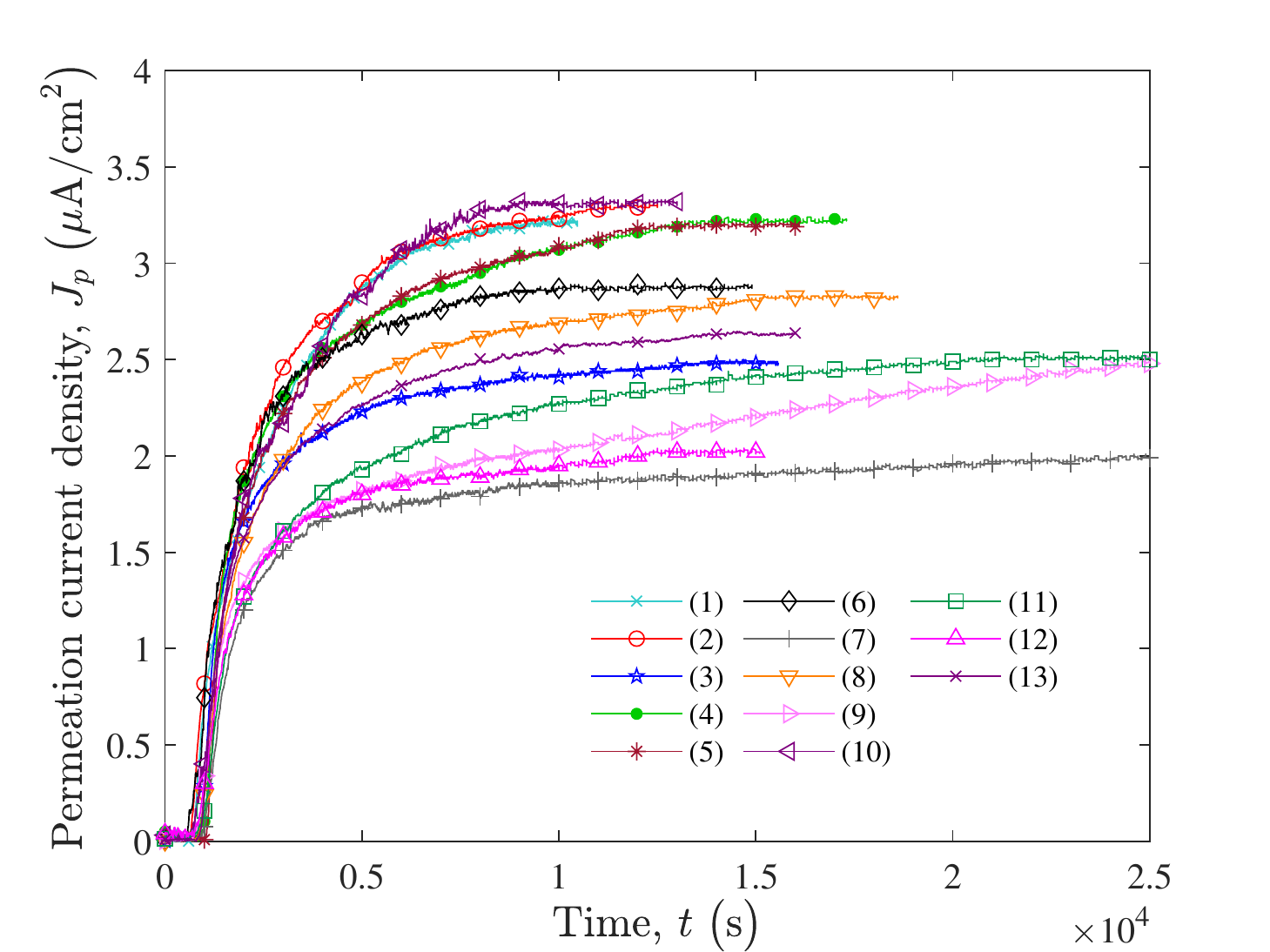}
        \caption{$1^{st}$ permeation transients ($J_{c}$ = 5 mA/cm$^2$) for cold-rolled pure Fe at 22$^\circ$C.}
        \label{fig:PermFirstExp}
\end{figure}

The $1^{st}$ permeation transients obtained for thirteen EP experiments performed on cold-rolled pure Fe at 22$^\circ$C are shown in Fig. \ref{fig:PermFirstExp}. Each experiment was stopped once the permeation current density reached steady state, $J_{ss1}$, which was defined as a variation in permeation current density of less than 0.05 $\mu$A/cm$^2$ over a 1000 second interval. The specimen thickness ($L$), steady state permeation current density ($J_{ss1}$), breakthrough time ($t_{bt}$), lag time ($t_{lag}$), and hydrogen diffusivities obtained by means of the breakthrough ($D_{bt1}$) and time lag ($D_{lag1}$) methods are shown for each transient in Table \ref{PermeationFirst}. The mean and standard deviation for each of these parameters are also provided in the table.

Despite the clear differences in the measured permeation transients, Table \ref{PermeationFirst} demonstrates that the average hydrogen diffusivity obtained by the breakthrough and lag time methods are very similar (5.72 and 6.09$\times10^{-11}$ m$^2$/s, respectively). This observation differs from a prior study conducted on this same material heat \cite{Zafra2022}, which reported a $\textgreater$30\% difference in the hydrogen diffusivity from the breakthrough and lag time approaches. However, it should be noted that this previous effort only performed two EP measurements and the present data demonstrates that an appreciable range of calculated diffusivities are possible for a given method (\textit{e.g.} 4.83 to 7.79$\times10^{-11}$ m$^2$/s for the lag time approach). Such results strongly suggest that large EP test matrices are necessary to obtain reliable hydrogen-metal interaction metrics.

\begin{table}[H]
 \centering
  \caption{$1^{st}$ permeation transient parameters for cold-rolled pure Fe at 22$^\circ$C.}
  \label{PermeationFirst}
\resizebox{\textwidth}{!}{
\begin{tabular}[t]{cccccccc}
\toprule
\text{Test} & \text{$L$ (mm)} & \text{$J_{ss1}$ ($\mu$A/cm$^2$)} & \text{$t_{bt1}$(s)} & \text{$D_{bt1}$ (m$^2$/s)} & \text{$t_{lag1}$(s)} & \text{$D_{lag1}$ (m$^2$/s)} \\
\midrule
(1) & 0.93 & 3.21 & 834 & 6.78$\times10^{-11}$ & 2596 & 5.55$\times10^{-11}$  \\
(2) & 0.90 & 3.30 & 776 & 6.76$\times10^{-11}$ & 2224 & 6.01$\times10^{-11}$  \\
(3) & 0.92 & 2.48 & 988 & 5.55$\times10^{-11}$ & 1795 & 7.79$\times10^{-11}$  \\
(4) & 0.93 & 3.22 & 1093 & 5.16$\times10^{-11}$ & 2310 & 6.23$\times10^{-11}$ \\
(5) & 0.92 & 3.20 & 1143 & 4.83$\times10^{-11}$ & 2504 & 5.62$\times10^{-11}$ \\
(6) & 0.88 & 2.87 & 753 & 6.69$\times10^{-11}$ & 1845 & 6.97$\times10^{-11}$  \\
(7) & 0.92 & 2.01 & 1087 & 5.13$\times10^{-11}$ & 2158 & 6.59$\times10^{-11}$ \\
(8) & 0.91 & 2.83 & 1025 & 5.33$\times10^{-11}$ & 2458 & 5.67$\times10^{-11}$ \\
(9) & 0.90 & 2.49 & 953 & 5.58$\times10^{-11}$ & 2808 & 4.83$\times10^{-11}$ \\
(10) & 0.93 & 3.31 & 945 & 5.99$\times10^{-11}$ & 2635 & 5.48$\times10^{-11}$ \\
(11) & 0.92 & 2.51 & 1088 & 5.04$\times10^{-11}$ & 2886 & 4.85$\times10^{-11}$ \\
(12) & 0.94 & 2.02 & 928 & 6.16$\times10^{-11}$ & 1998 & 7.30$\times10^{-11}$  \\
(13) & 0.91 & 2.63 & 1007 & 5.40$\times10^{-11}$ & 2203 & 6.29$\times10^{-11}$ \\
\midrule
Average & 0.92 & 2.78 & 971 & 5.72$\times10^{-11}$ & 2340 & 6.09$\times10^{-11}$  \\
SD & 0.02 & 0.46 & 123 & 0.68$\times10^{-11}$ & 346 & 0.90$\times10^{-11}$ \\
\bottomrule
\end{tabular}}
\end{table}

\begin{figure}[H]
     \centering
         \centering
         \includegraphics[width=0.8\textwidth]{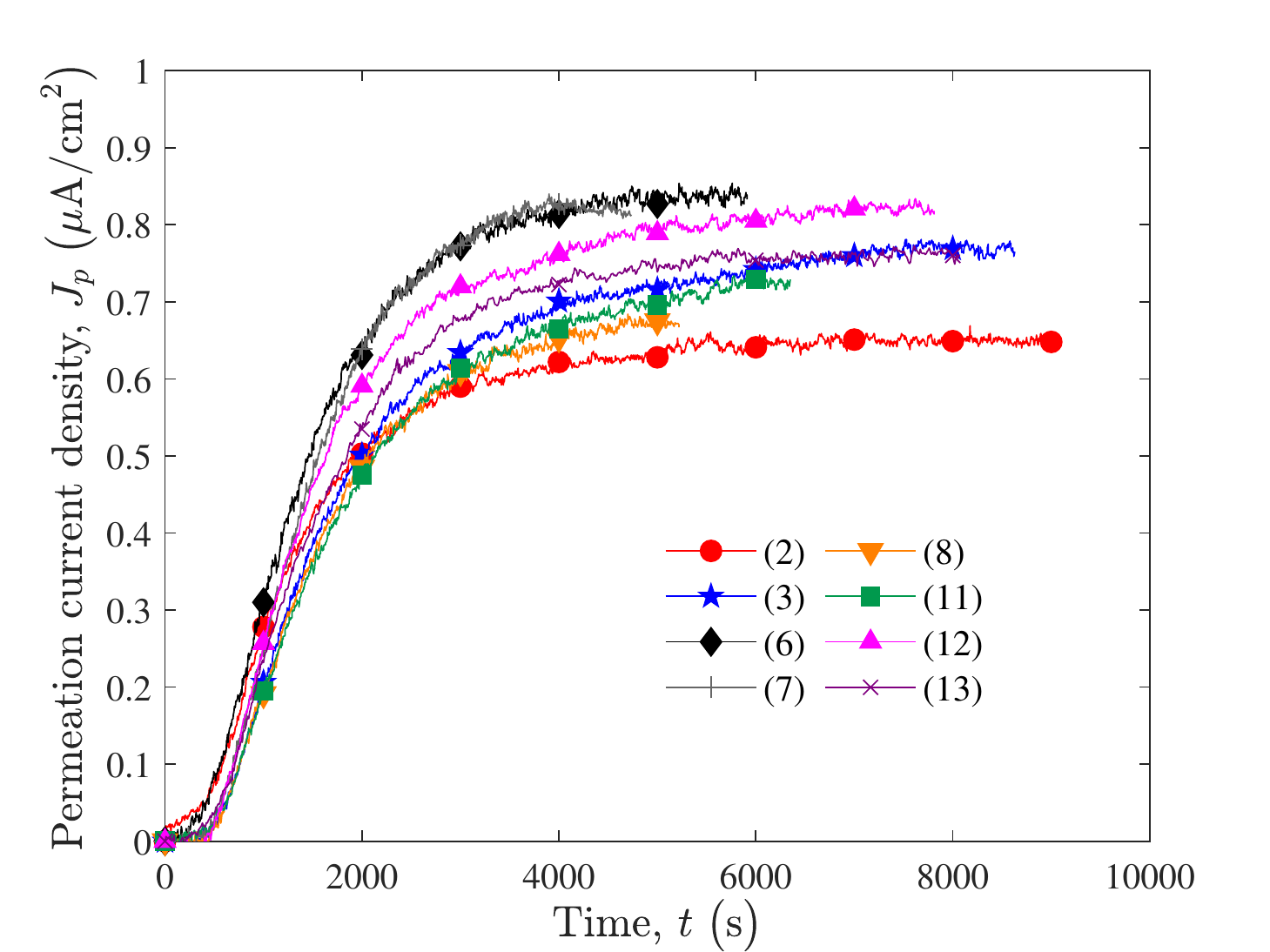}
        \caption{$2^{nd}$ permeation transients ($J_{c}$ = 10 mA/cm$^2$) for cold-rolled pure Fe at 22$^\circ$C.}
        \label{fig:PermSecExp}
\end{figure}

After the completion of the $1^{st}$ permeation transient, a $2^{nd}$ permeation transient was performed on eight of the experiments by increasing $J_c$ to 10 mA/cm$^2$. As both hydrogen trapping \cite{Fallahmohammadi2013} and unsteady surface conditions \cite{Turnbull1994} will be attenuated during the $2^{nd}$ transient, it is expected that the scatter in the calculated hydrogen diffusivity should be reduced relative to the $1^{st}$ transient. The curves corresponding to the $2^{nd}$ transient are shown in Fig. \ref{fig:PermSecExp}, which demonstrates that each experiment reached a stable permeation current density ($J_{ss2}$), within 10000 seconds. $J_{ss2}$, along with $t_{bt2}$, $t_{lag2}$ and the associated hydrogen diffusivities ($D_{bt2}$ and $D_{lag2}$) are reported in Table \ref{PermeationSecond}. The mean and standard deviation for each of these parameters are also included in the table. As noted for the $1^{st}$ permeation transient, the average hydrogen diffusivities obtained for the $2^{nd}$ transient using the breakthrough and lag time methods are very similar, 8.64 and 8.22$\times10^{-11}$ m$^2$/s respectively. However, these values are approximately 1.5 times higher than the diffusivities that were calculate from the $1^{st}$ permeation transient (5.72-6.09$\times10^{-11}$ m$^2$/s), consistent with a reduced contribution of hydrogen trapping \cite{Zhang1998,Fallahmohammadi2013,Svoboda2014,Zafra2020b}. 

\begin{table}[H]
 \centering
  \caption{$2^{nd}$ permeation transient parameters for cold-rolled pure Fe at 22$^\circ$C.}
  \label{PermeationSecond}
\resizebox{\textwidth}{!}{
\begin{tabular}[t]{cccccccc}
\toprule
\text{Test} & \text{$L$ (mm)}&\text{$J_{ss2}$ ($\mu$A/cm$^2$)} & \text{$t_{bt2}$(s)} & \text{$D_{bt2}$ (m$^2$/s)} & \text{$t_{lag2}$(s)} & \text{$D_{lag2}$ (m$^2$/s)} \\
\midrule
(2) & 0.90 & 0.65 & 525 & 9.99$\times10^{-11}$ & 1473 & 9.08$\times10^{-11}$ \\
(3) & 0.92 & 0.77 & 702 & 7.81$\times10^{-11}$ & 1892 & 7.39$\times10^{-11}$ \\
(6) & 0.88 & 0.84 & 518 & 9.73$\times10^{-11}$ & 1514 & 8.49$\times10^{-11}$ \\
(7) & 0.92 & 0.82 & 687 & 8.12$\times10^{-11}$ & 1660 & 8.57$\times10^{-11}$ \\
(8) & 0.91 & 0.67 & 680 & 8.03$\times10^{-11}$ & 1715 & 8.12$\times10^{-11}$ \\
(11) & 0.92 & 0.72 & 682 & 8.05$\times10^{-11}$ & 1880 & 7.44$\times10^{-11}$ \\
(12) & 0.94 & 0.82 & 640 & 8.93$\times10^{-11}$ & 1681 & 8.67$\times10^{-11}$ \\
(13) & 0.91 & 0.76 & 640 & 8.49$\times10^{-11}$ & 1730 & 8.01$\times10^{-11}$ \\
\midrule
Average & 0.91 & 0.76 & 634 & 8.64$\times10^{-11}$ & 1693 & 8.22$\times10^{-11}$ \\
SD & 0.02 & 0.07 & 73 & 0.83$\times10^{-11}$ & 150 & 0.60$\times10^{-11}$ \\
\bottomrule
\end{tabular}}
\end{table}

\subsection{Isothermal TDS tests}

\begin{figure}[H]
         \stackinset{r}{-.15\textwidth}{t}{-.01\textwidth}
  {\includegraphics[width=0.6\textwidth]{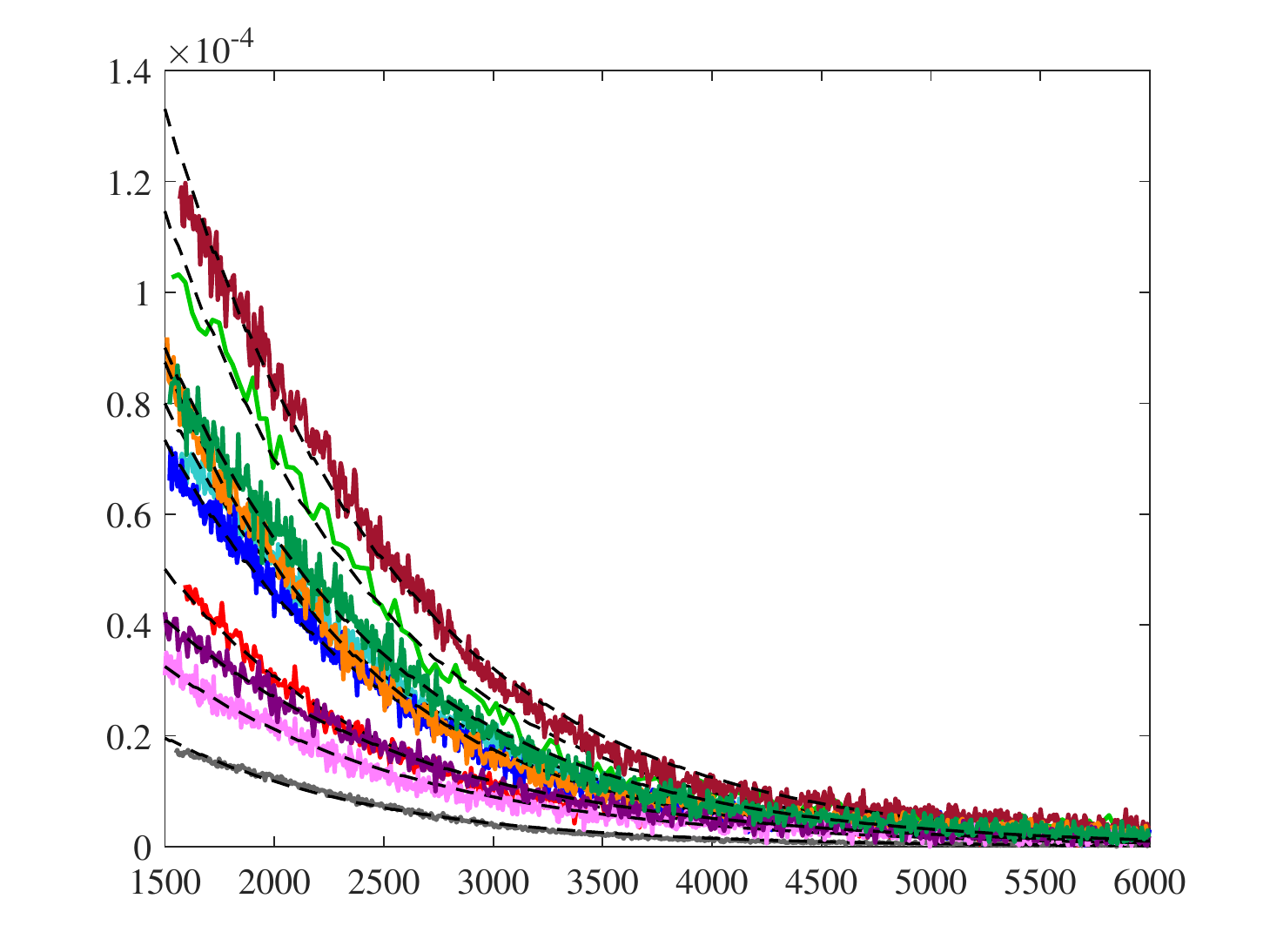}}
  {\includegraphics[width=0.85\textwidth]{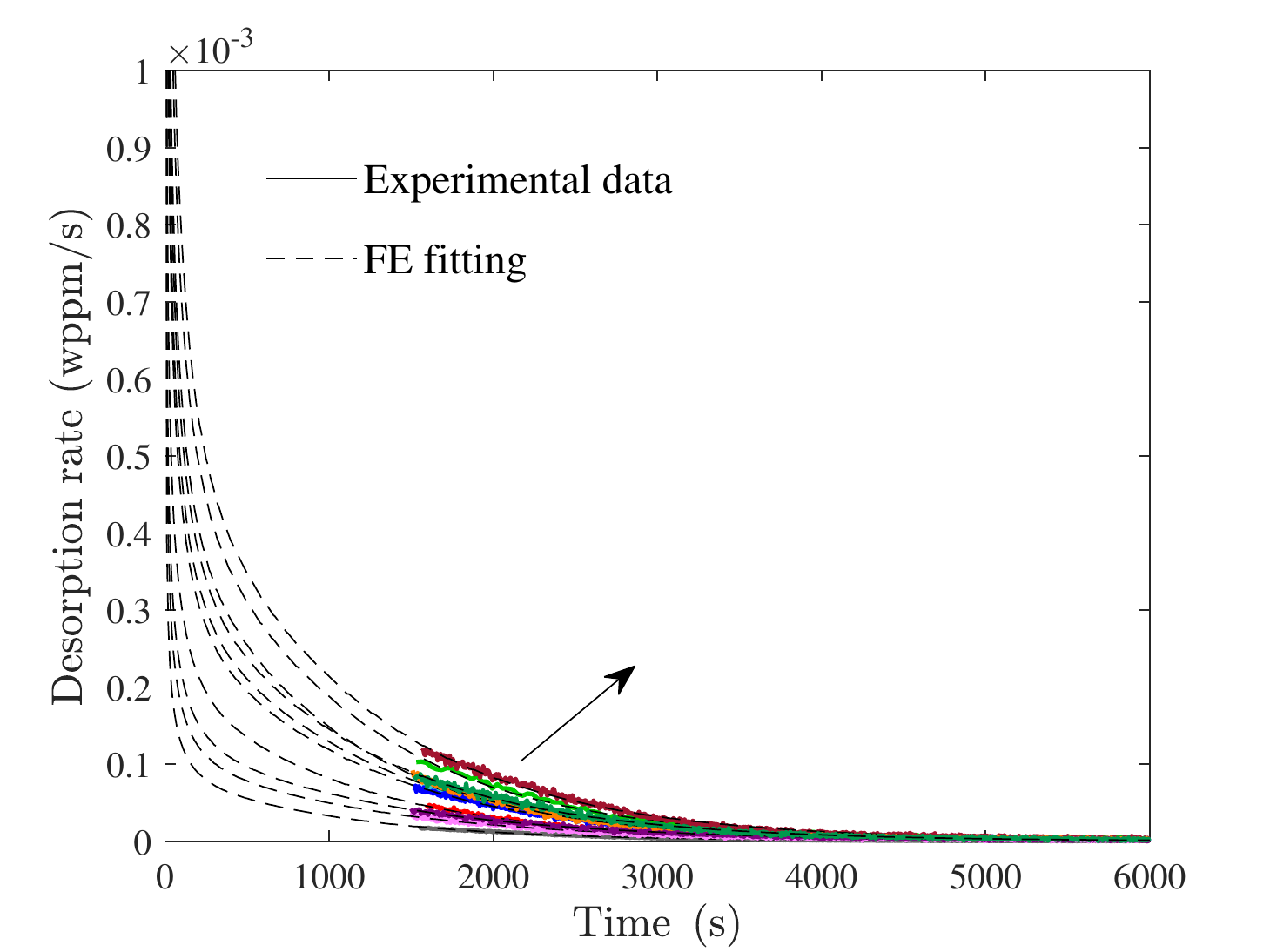}}
        \caption{Experimental hydrogen desorption rate vs. time profiles obtained from ITDS experiments performed at 22$^\circ$C (solid lines) and the best fit determined from finite element analysis (dashed lines).}
        \label{fig:IsoDesorption}
\end{figure}

The hydrogen desorption rate versus time profiles measured during ten ITDS experiments conducted at 22$^\circ$C are shown in Fig. \ref{fig:IsoDesorption}. The best fit to Fick's second law, determined from finite element (FE) simulations, is indicated by the corresponding dashed lines. For all cases, the FE calculations closely capture the experimental behaviour. The initial diffusible hydrogen concentration ($C_{0iso}$) and diffusion coefficient ($D_{iso}$) that yielded the best fit of the experimental curve, as well as the coefficient of determination, R$^2$, which provides a goodness-of-fit measure, are reported in Table \ref{Isothermal}. $C_{0iso}$ represents the area below the ITDS profile starting from $t=0$. The mean and standard deviation for these two parameters are provided at the bottom of the table. Two observations can be made from these data. First, the standard deviation for $D_{iso}$ (0.36$\times10^{-11}$ m$^2$/s) is generally half of the standard deviation observed across the permeation approaches. Second, the average  $D_{iso}$ lies in between the diffusivities observed for the first and second permeation transients, but notably closer to the second transient result.

\begin{table}[H]
 \centering
  \caption{Initial diffusible hydrogen concentration, $C_{0iso}$, and diffusivity values, $D_{iso}$, determined from ITDS experiments.}
  \label{Isothermal}
\begin{tabular}[t]{ccccc}
\toprule
\text{Test} & \text{$L$ (mm)} & \text{$C_{0iso}$ (wppm)} & \text{$D_{iso}$ (m$^2$/s)} & $R^2$ \\
\midrule
(1) & 0.87 & 0.43 & 7.27$\times10^{-11}$ & 0.988 \\
(2) & 0.90 & 0.27 & 7.98$\times10^{-11}$ & 0.992 \\
(3) & 0.92 & 0.40 & 8.20$\times10^{-11}$ & 0.985 \\
(4) & 0.90 & 0.63 & 8.02$\times10^{-11}$ & 0.991 \\
(5) & 0.90 & 0.71 & 7.72$\times10^{-11}$ & 0.994 \\
(6) & 0.88 & 0.84 & 8.00$\times10^{-11}$ & 0.985 \\
(7) & 0.86 & 0.50 & 7.94$\times10^{-11}$ & 0.989 \\
(8) & 0.92 & 0.17 & 7.28$\times10^{-11}$ & 0.990 \\
(9) & 0.93 & 0.21 & 7.23$\times10^{-11}$ & 0.993 \\
(10) & 0.91 & 0.48 & 7.97$\times10^{-11}$ & 0.986 \\
\midrule
Average & 0.90 & 0.46 & 7.76$\times10^{-11}$ & 0.989 \\
SD & 0.02 & 0.22 & 0.36$\times10^{-11}$ & 0.003 \\
\bottomrule
\end{tabular}
\end{table}

\section{Discussion}
\label{Sec:Discussion}

The preceding results provide a comprehensive dataset that can be leveraged to compare the relative efficacy of EP and ITDS methods for determining hydrogen diffusivity. The objectives of the following discussion are: (1) comment on the diffusivities calculated amongst the tested approaches, (2) identify the factors likely responsible for the test-to-test scatter that exists in each technique, and (3) discuss the implications of these results in the context of the optimal approach for assessing hydrogen diffusivity.

\subsection{Comparison of diffusivities obtained in EP and ITDS experiments}
\label{Subsec:ComparisonDiff}
\begin{figure}[H]
     \centering
         \centering
         \includegraphics[width=0.85\textwidth]{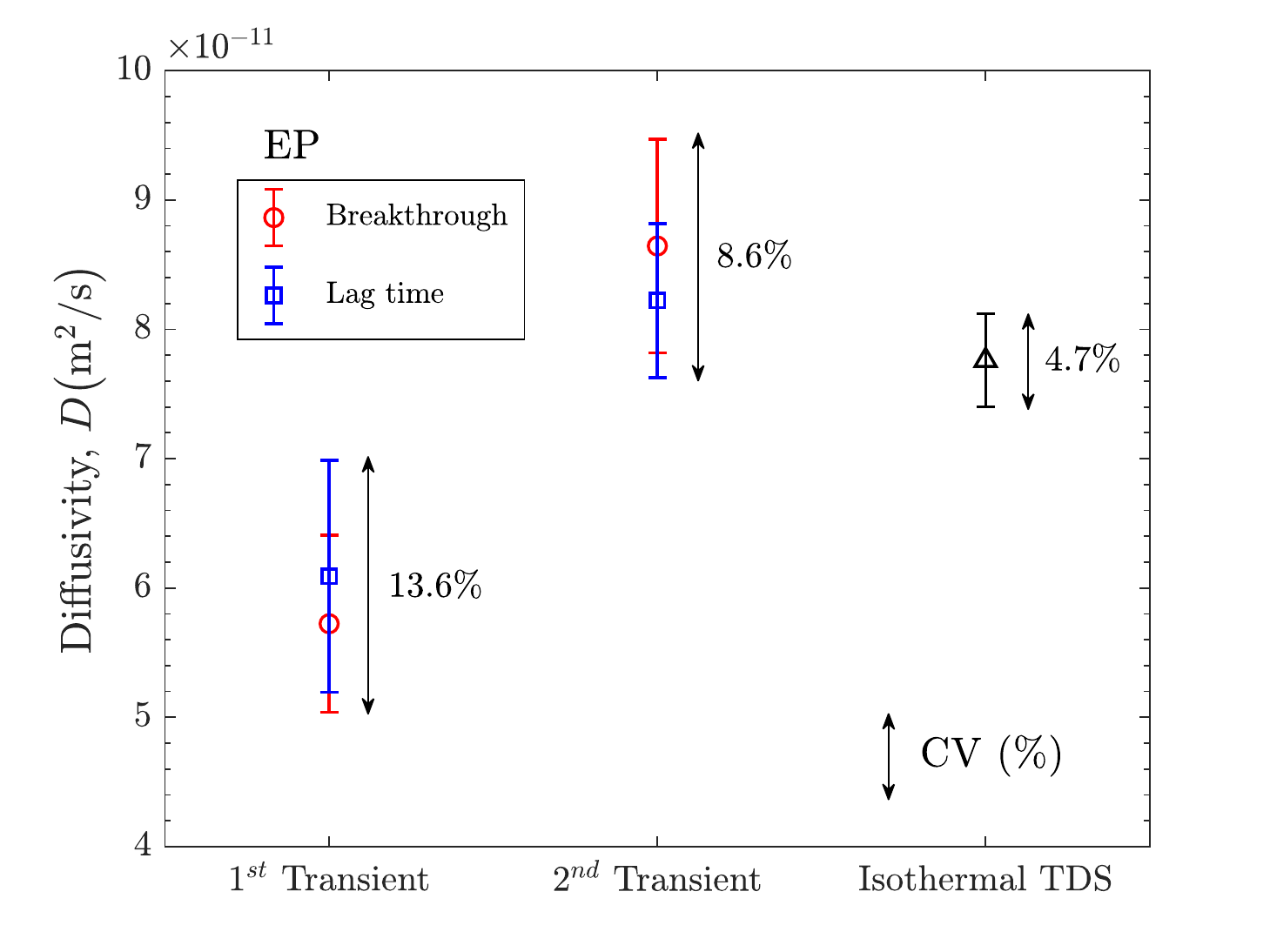}
        \caption{Average diffusivity and standard deviation (represented by the error bars) obtained from electopermeation ($1^{st}$ and $2^{nd}$ transients) and ITDS experiments. The coefficient of variation CV(\%), indicated by the black arrows, is also included for each group.}
        \label{fig:ComparisonScattering}
\end{figure}

Fig. \ref{fig:ComparisonScattering} shows the average diffusivity and standard deviation (represented by the error bars) obtained from the $1^{st}$ and $2^{nd}$ transients of the permeation experiments (breakthrough and lag time methods) as well as the ITDS experiments. The coefficient of variation (CV), which is defined as the ratio of the standard deviation and the mean, is also provided to quantify the dispersion in the measured hydrogen diffusivity for each approach, and is represented by black arrows in Fig. \ref{fig:ComparisonScattering}. 

Three observations are notable from the summary presented in Fig. \ref{fig:ComparisonScattering}. First, for a given permeation transient, the lag time and breakthrough time methods yield nominally identical results, with the average of one approach always well within the error bars of the other. Second, ITDS yields a similar average hydrogen diffusivity as the $2^{nd}$ permeation transient ($\sim$7.8 v. $\sim$8.4x10$^{-11}$ m$^2$/s), with both being increased relative to the diffusivity measured during the $1^{st}$ permeation transient ($\sim$6x10$^{-11}$ m$^2$/s). Lastly, both the standard deviation and CV are significantly reduced for the ITDS experiments as compared to the permeation experiments, with the CV being nearly half and one-third of that observed for the $2^{nd}$ and $1^{st}$ permeation transients, respectively.

Considering the first observation, recall that the lag time and breakthrough time approaches represent closed-form solutions to Fick's 2$^{nd}$ law for the same boundary conditions. In other words, these methods represent two points along the same normalized flux ($J_p$/$J_{ss}$) versus time curve. As long as the assumed boundary conditions for these solutions are generally reasonable for the performed experiment, it follows that the two approaches should yield similar \textit{average} diffusivity values. Regarding the second observation, the improved agreement in the average hydrogen diffusivities measured using the $2^{nd}$ permeation transient and ITDS methods relative to the $1^{st}$ transient is consistent with expectations for the boundary conditions of each experiments. Specifically, while the $1^{st}$ permeation transient begins with a nominally hydrogen-free membrane, the second permeation transient and ITDS experiments both start with a nominally uniform, non-zero hydrogen content across the specimen thickness. As such, trapping effects should be attenuated during the $2^{nd}$ permeation transient and ITDS experiments, resulting in an increased hydrogen diffusivity relative to the $1^{st}$ transient. 

Lastly, the most striking observation from the present data in Fig. \ref{fig:ComparisonScattering} is the reduced scatter in the ITDS-measured diffusivity as compared to the permeation-based measurements. While prior work reported a lower error for ITDS-based measurements \cite{Zafra2022}, this previous study only performed a small number of experiments. The current work confirms the increased precision of ITDS relative to EP for a much larger experimental matrix ($\geqslant$ 10 of each kind). Critically, the present results also highlight the propensity for increased error in EP experiments relative to ITDS, regardless of employed analysis approach or the use of multiple transients. Interestingly, the difference in the CV between the $1^{st}$ (13.6\%) and $2^{nd}$ (8.6\%) transients suggests that only a fraction of error in EP experiments is likely related to trapping influences, suggesting that other factors provide a more significant contribution to the error in EP experiments.

\subsection{Identification of the sources of error}
\subsubsection{Sources of error in EP tests}

Several authors have previously outlined the factors that may affect the repeatability of EP measurements \cite{Gonzalez1969,Zakroczymski1982,Kiuchi1983,Turnbull1995}. These prior efforts have collectively established that hydrogen uptake and transport in metals during EP tests is sensitive to the following broad categories:

\begin{enumerate}[label=(\roman*)]
    \item Solution composition/pH and temperature. These factors can influence the amount of absorbed hydrogen as well as trap occupancy, which will modify the hydrogen diffusivity. While variations in these parameters would induce scattering in EP experiments, they are not considered a likely source of error in the current experiments due to the significant efforts made to control these variables.
    
    \item Trapping phenomena. The rate of transport of hydrogen atoms through the membrane is affected by both reversible and irreversible trapping \cite{Fallahmohammadi2013,Caskey1975}. Variability in this effect could manifest in two ways: (1) sample-to-sample variations in trap site characteristics and (2) evolution in the efficacy of trapping as the permeation experiment progresses and traps are filled. While the former cannot be rigorously excluded, all samples were prepared from the same plate of Fe using the same approaches to minimize such effects. An effect of the latter is clearly demonstrated by the results in Fig. \ref{fig:ComparisonScattering}, where the diffusivity during the $2^{nd}$ transient was notably increased relative to that measured in the $1^{st}$ transient. However, the fact that the $2^{nd}$ transient still exhibits elevated levels of scatter suggests that trapping is not the primary source of the observed variations in EP data.
    
    \item Surface effects. The specimen's surface condition and thickness play an essential role in minimizing surface effects during hydrogen permeation through a metal membrane. Typically, a well-polished surface \cite{Lopez-Suarez2010}, the application of surface catalysts such as palladium coatings \cite{Makhlouf1991,Zhang1999} or the use of hydrogen recombination poisons \cite{Zakroczymski1976,Juang1994} are employed to minimize surface effects on hydrogen dissociation and subsequent uptake into the membrane. Additionally, it is important that the sample be sufficiently thick such that these surface effects do not dominate over bulk diffusivity. Prior work demonstrated that surface effects in pure Fe are minimal for samples with thicknesses larger than 0.5 mm \cite{Kittel2016}. While this suggests that surface effects are not the likely source of scatter in the current experiments, the influence of subtle differences in Pd layer thickness on the oxidation side of the cell and surface finish on the reduction side cannot be completely neglected. A possible role of surface effects on the scatter would also be consistent with postulated sources of error for permeation measurements in the literature \cite{Gonzalez1969,Kumnick1974,Boes1976}.
    
    \item Boundary conditions at the entry and the exit surfaces. The use of galvanostatic or potentiostatic charging can modify the boundary conditions on the reduction side of the EP experiment and therefore determine the most appropriate method of analysis of the permeation transient \cite{Pumphrey1980,Montella1999,Zhang1999}. Assuming incorrect boundary conditions can have a significant impact on the average diffusivity \cite{Montella1999}, while scatter can also be induced by failing to fully meet the assumed boundary conditions (\textit{i.e.}, due to a passive film) or through the evolution of the boundary conditions during an EP experiment. 
    
\end{enumerate}

While trapping influences cannot be completely ruled out, the above review indicates that assumed boundary conditions (and the role of surface effects on those boundary conditions) likely provides a more significant contribution to the increased error observed in the EP measurements. In the following section, the importance and influence of assumed boundary conditions on the calculated hydrogen diffusivity is assessed through the application of three models for EP.

\subsubsection{Influence of assumed boundary conditions for analysis of permeation data}

In the majority of the literature, the hydrogen diffusivity is determined from EP data using the breakthrough and lag time methods \cite{Sun2015}. These approaches are employed due to their closed-form nature, which avoids the need for more complex fitting algorithms. However, it is important to note that these methods represent specific solutions (at $J_{p}$/$J_{ss}$=0.1 and 0.63, respectively) to Fick's second law for 1-D permeation under the boundary conditions of a constant hydrogen concentration of some finite amount at the membrane's entry side ($C_{x=0}$=$C_{0app}$) and of zero at the exit side ($C_{x=L}=0$). Under such constant concentration (CC) conditions, the complete normalized permeation flux as a function of time relationship is described by:
 \begin{equation}\label{eq:FitCC}
    \frac{J_p}{J_{ss}}= 1+2\sum_{n=1}^\infty(-1)^n\exp\left(-\frac{n^2\pi^2D_{CC}t}{L^2}\right)
    \end{equation}
    
\noindent where $L$ is the specimen thickness and $D_{CC}$ is the diffusivity for these assumed boundary conditions. Critically, this model (and therefore the lag time and breakthrough time solutions) is only rigorously applicable to EP experiments conducted using potentiostatic conditions on the reduction side of the permeation cell \cite{Zhang1999,Montella1999}. This arises from the fact that the hydrogen fugacity, and therefore concentration, is directly related to the hydrogen overpotential, which is established by the applied electrochemical potential and solution composition \cite{Subramanyan1981}. This requirement can be functionally met using galvanostatic charging so long as the surface conditions and environment do not change \cite{Turnbull1995}. Such a scenario results in a nominally constant applied potential on the specimen surface, but careful monitoring is required to confirm compliance with this boundary condition.

While it is possible to apply the CC model to EP experiments conducted under galvanostatic charging conditions, it is most rigorous to evaluate such experiments with a constant flux (CF) model. This model assumes 1-D permeation under a constant hydrogen flux at the entry side of the membrane ($J_{x=0}$=-F$J_c$) and a hydrogen concentration of zero at the exit surface ($C_{x=L}$=0) \cite{Pumphrey1980,Zhang1999}:
 \begin{equation}\label{eq:FitCF}
    \frac{J_p}{J_{ss}}= 1-\frac{4}{\pi}\sum_{n=0}^\infty \frac{(-1)^n}{2n+1}\exp\left(-\frac{(2n+1)^2\pi^2D_{CF}t}{4L^2}\right)
    \end{equation}

While the CC and CF models represent idealized conditions that do not consider surface effects \cite{Montella1999}, hydrogen uptake in a real system will depend on the kinetics of the following absorption-desorption reaction:
    \begin{equation}\label{eq:SurfReactions}
    \ce{ H_{adsorbed} <=>[\ce{K_{abs}}][\ce{K_{des}}] H_{absorbed}} 
    \end{equation}
Several authors have attempted to account for these surface effects through the development of more complex permeation models \cite{Pumphrey1980,Montella1999,Wang1936,Zhang1999}. For example, the flux continuity (FC) model was first proposed by Wang \cite{Wang1936} to describe hydrogen gas permeation while taking into account absorption and desorption reactions. Zhang \textit{et al.} \cite{Zhang1999} simplified the absorption-desorption reactions used in the FC model and applied several assumptions (e.g., similar surface conditions at the entry and exit surfaces, and constant surface coverage ($\theta$) with permeation time), which allowed the extension of this model to describe EP experiments. The normalized permeation flux as a function of time for these assumptions is then expressed as: 
 \begin{equation}\label{eq:FitFC}
    \frac{J_p}{J_{ss}}= 1+2(2+\xi)\sum_{n=1}^\infty \frac{\xi cos(\eta_n)-\eta_n sin(\eta_n)}{\eta_n^2+\xi^2+2\xi}\exp\left(-\frac{(\eta_n)^2 D_{FC} t}{L^2}\right)
    \end{equation}
 
\noindent where $\xi$ is the ratio between the desorption constant $k_{des}$ and the diffusion constant $D/L$ in the metal, and the dimensionless parameter $\eta_n$ is the $n$th positive root of: 
\begin{equation}\label{eq:Roots}
   tan(\eta_n)= \frac{2 \xi \eta_n}{\eta_n^2-\xi^2}
    \end{equation}

As the desorption constant, $k_{des}$, is analogous to a surface-limited mass transfer coefficient \cite{Makhlouf1991}, a lower $k_{des}$ implies an increased effect of the surface on the overall mass transport. Therefore, the $\xi$ parameter can be used as a proxy for the relative influence of surface effects on the permeation transient, where a reduction in $\xi$ similarly indicates an increased role of hydrogen absorption and desorption during permeation. In addition to explicitly incorporating surface influences, the FC model provides a more general solution to the permeation problem, which under specific conditions result in the FC modeling becoming identical to either the CC or CF models. For example, if the value of $\xi$ becomes sufficiently large or small, then the FC model reduces to the CC and CF models, respectively \cite{Montella1999,Zhang1999}. However, this generality comes at the cost of increased analysis complexity relative to the CC and CF models since both the diffusivity, $D_{FC}$, and $\xi$ must be numerically determined.

\begin{figure}[htp]
     \centering
    \begin{subfigure}{0.5\textwidth}
         \centering
         \includegraphics[width=\textwidth]{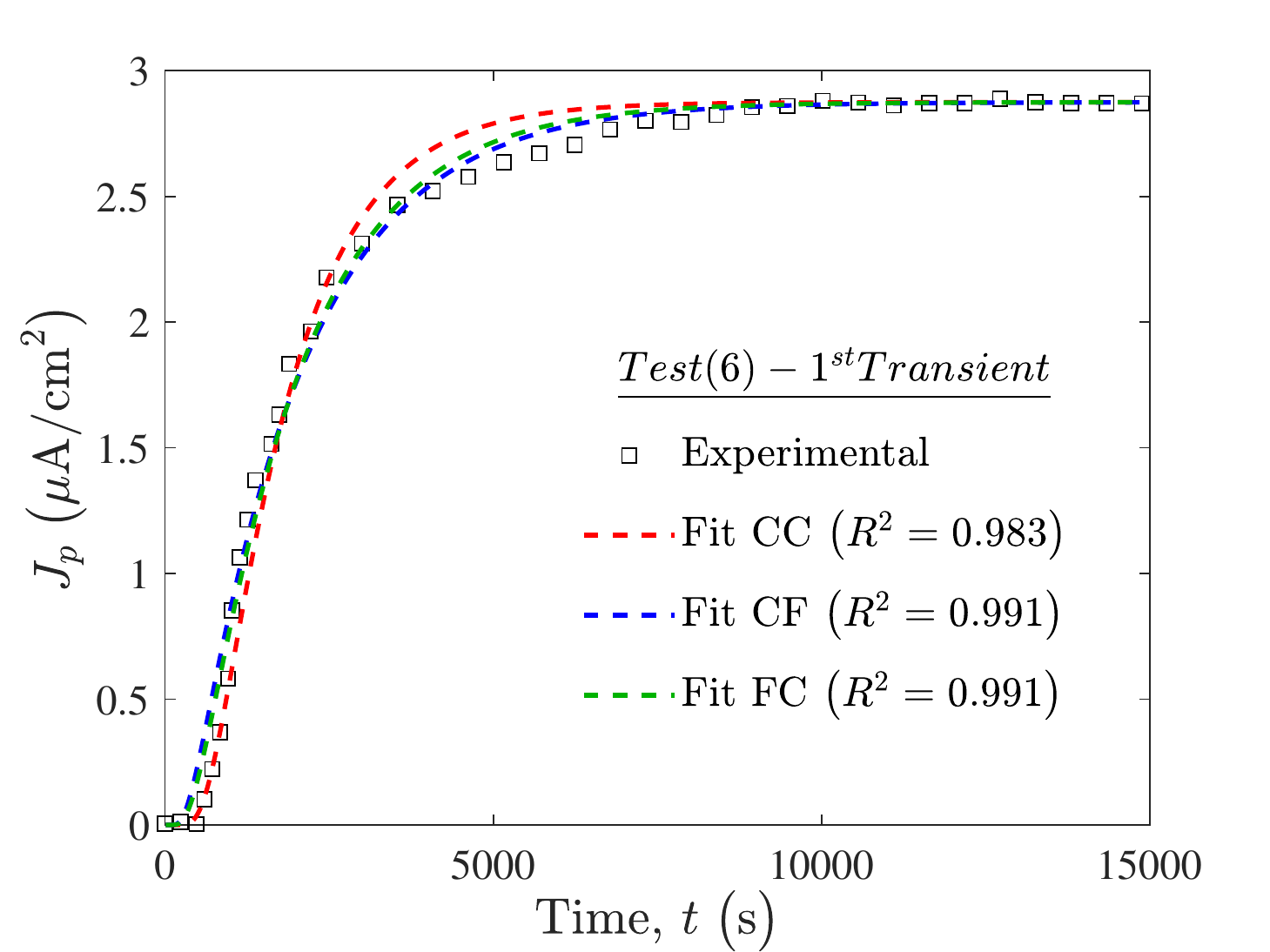}
            \caption{}
         \end{subfigure}\hfill
    \begin{subfigure}{0.5\textwidth}
         \centering
         \includegraphics[width=\textwidth]{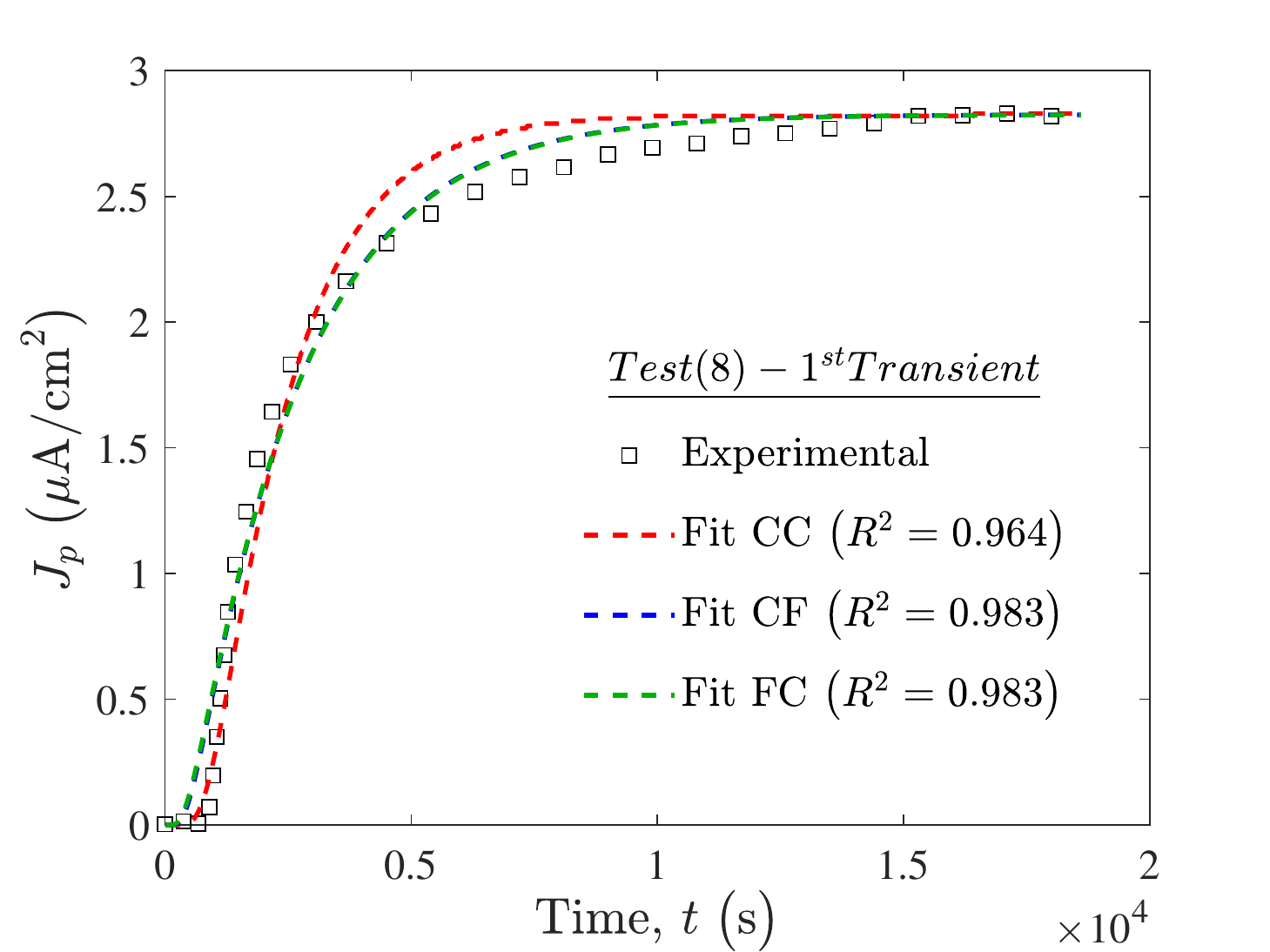}
            \caption{}
         \end{subfigure}
    \begin{subfigure}{0.5\textwidth}
         \centering
         \includegraphics[width=\textwidth]{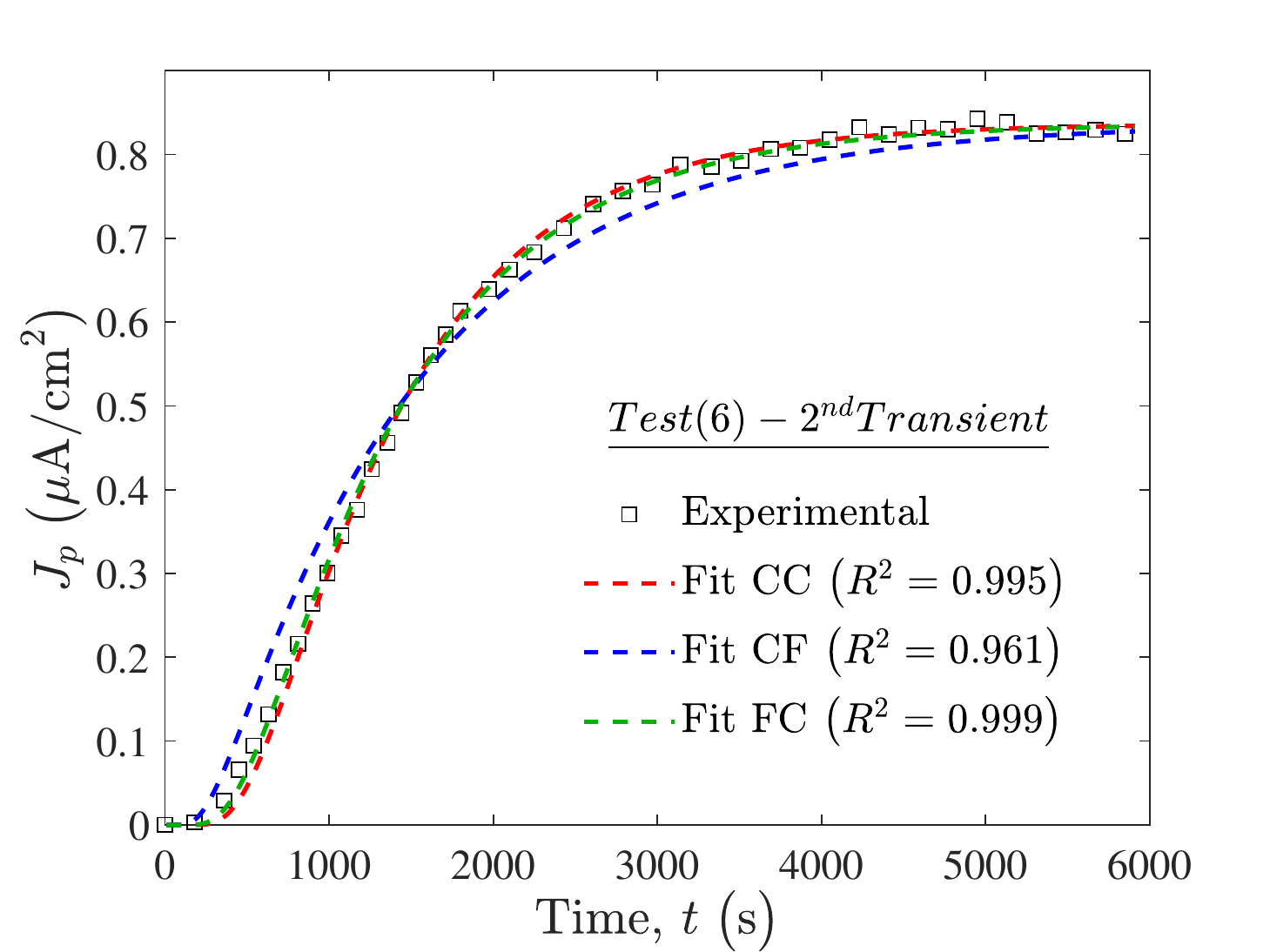}
            \caption{}
         \end{subfigure}\hfill
    \begin{subfigure}{0.5\textwidth}
         \centering
         \includegraphics[width=\textwidth]{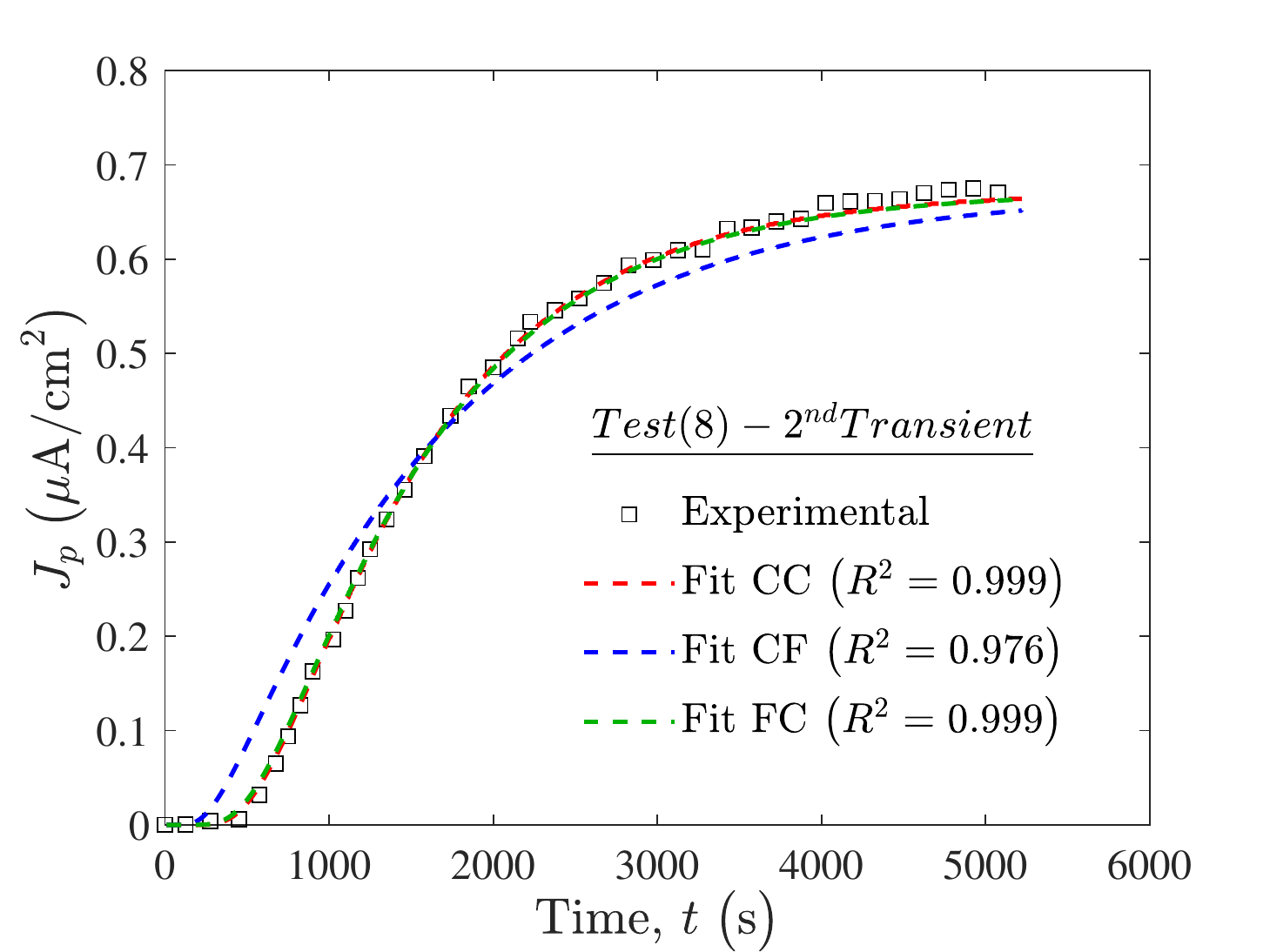}
            \caption{}
         \end{subfigure}\hfill
        \caption{Comparison of the CC, CF, and FC model fits to two representative (a, b) 1$^{st}$ and (b, c) 2$^{nd}$ transient EP experiments.}
        \label{fig:PermeFits}
\end{figure}

To assess the effect of assumed boundary conditions, every permeation transient was iteratively fit to each model using Matlab with the diffusivity being the free parameter for the CC and CF model, while both the diffusivity and $\xi$ parameter were fit for the FC model. As shown in Fig. \ref{fig:PermeFits} for two representative 1$^{st}$ and 2$^{nd}$ transients, all three models reasonably describe each permeation transient as quantified by the R$^2$ coefficients listed in the figure, though subtle differences can be identifed. For example, the CF and FC models yield the best (and nominally identical) fits to the first permeation transients (Fig. \ref{fig:PermeFits}a-b); the good agreement of the CF model is consistent with expectations given the use of galvanostatic charging in the current experiments. Conversely, the 2$^{nd}$ permeation transients (Fig. \ref{fig:PermeFits}c-d) are best fit by the CC model (which is identical for this case to the FC model). While the 2$^{nd}$ transient was also conducted under galvanostatic conditions, these results indicate that the surface was sufficiently conditioned to enable a CC-like boundary condition \cite{Turnbull1995}.

\begin{figure}[H]
     \centering
         \centering
         \includegraphics[width=0.85\textwidth]{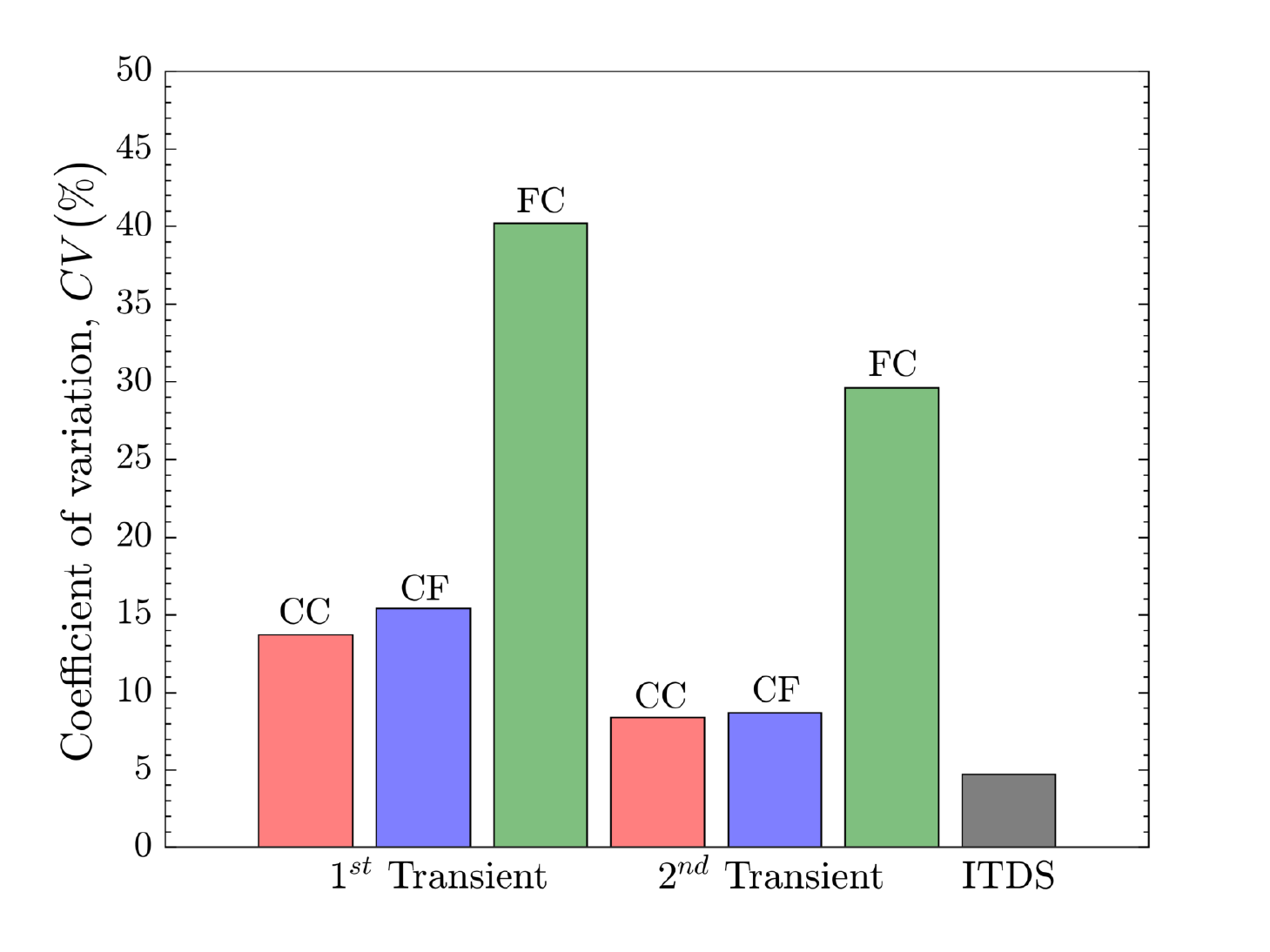}
        \caption{Coefficient of variation, CV, associated to each method.}
        \label{fig:ComparisonMethods}
\end{figure}

The diffusivities $D_{CC}$, $D_{CF}$, $D_{FC}$, and $\xi$ determined for each 1$^{st}$ and 2$^{nd}$ transient measurement, along with the R$^2$ for each fitting are provided in the Supplementary Material. The mean and standard deviation for these parameters are listed in Table \ref{PermeationFitBC}, while a barplot of the CV for each model's calculated diffusivity is shown in Fig. \ref{fig:ComparisonMethods}. These fitted parameters reveal two key observations. First, the calculated hydrogen diffusivity is strongly dependent on the applied model, with nearly five-fold and four-fold difference observed for the 1$^{st}$ and 2$^{nd}$ transients, respectively. Critically, these differences exist despite the models exhibiting similar goodness of fit to the experimental data (e.g., the CF and FC model during the 1$^{st}$ transient and the CC and FC model during the 2$^{nd}$ transient). Second, the choice of boundary conditions also has an influence on the observed scatter of permeation experiments. For example, the FC model has an approximately 3-fold larger CV (reaching 40\% for the 1$^{st}$ transient) relative to the CC and CF models, which had generally similar CV values. Such increased variability for the FC model is likely driven by the need to fit two independent parameters ($D_{FC}$ and $\xi$). Lastly, it is interesting to note that the lowest error observed across the full fit approaches is ~8\%. Zhang \textit{et al.} \cite{Zhang1998} reported a similar error ($\sim$11\%) for a large ($\sim$5 specimen) EP test matrix that was analyzed with full-fitting methods, while Svoboda and co-workers \cite{Svoboda2014} observed an 8\% error when analyzing a large number of duplicate EP experimental pairs via full-fitting. Such data suggests that an error of 8-10\% may be the lower-bound error for the EP method.

\begin{table}[H]
 \centering
  \caption{Diffusivity values obtained in the $1^{st}$ and $2^{nd}$ transients of all the permeation experiments by fitting the CC, CF and FC models.}
  \label{PermeationFitBC}
\resizebox{\textwidth}{!}{
\begin{tabular}[t]{ccccccccc}
\toprule
\multirow{2}{*}{Transient} & \multirow{2}{*}{Value} & \multicolumn{2}{c}{Constant Concentration} & \multicolumn{2}{c}{Constant Flux} & \multicolumn{3}{c}{Flux Continuity}\\
& & \text{$D_{CC}$ (m$^2$/s)} & \text{R$^2$} & \text{$D_{CF}$ (m$^2$/s)} & \text{R$^2$} & \text{$D_{FC}$ (m$^2$/s)} & $\xi$ & \text{R$^2$} \\
\midrule
\multirow{2}{*}{$1^{st}$} & Average & 5.76$\times10^{-11}$ & 0.938 & 1.60$\times10^{-10}$ & 0.963 & 2.53$\times10^{-10}$ & 1.09 & 0.965 \\
& SD & 0.79$\times10^{-11}$ & 0.09 & 0.25$\times10^{-10}$ & 0.06 & 1.03$\times10^{-10}$ & 0.49 & 0.04 \\
\midrule
\multirow{2}{*}{$2^{nd}$} & Average & 8.42$\times10^{-11}$ & 0.996 & 2.44$\times10^{-10}$ & 0.985 & 1.49$\times10^{-10}$ & 7.0 & 0.998 \\
& SD & 0.71$\times10^{-11}$ & 0.003 & 0.21$\times10^{-10}$ & 0.01 & 0.44$\times10^{-10}$ & 4.8 & 0.002\\
\bottomrule
\end{tabular}}
\end{table}

The preceding analysis underscores the difficulty of rigorously interpreting EP data, since the applicability of a given model is affected by the employed electrochemical boundary conditions (which may evolve during an experiment) and whether or not to consider absorption/desorption kinetics. While uncertainties associated with choosing the best model to fit experimental data always exist, EP fitting appears unique in that similarly well-fit models (e.g., the CC and FC model fits to the 2$^{nd}$ EP transient) yield two-to-three fold differences in best-fit diffusivity. In other words, the choice of model itself for interpreting EP data becomes a source of error for EP experiments. This observation is consistent with the conclusions of other authors \cite{Montella1999,Pound1993,Zhang1999}, who have each suggested that the primary challenge for EP methods is ensuring the theoretical conditions assumed EP data analysis are valid for the employed experimental conditions. Critically, these challenges are magnified by the widespread use of simple closed-form solutions (e.g., the lag and breakthrough time methods). Such approaches are convenient, but do not provide feedback on the overall goodness of the chosen model's fit and can lead to potential misinterpretations. For example, variations between lag time and breakthrough time calculations often are attributed to trapping effects, which undoubtedly affect EP measurements \cite{Caskey1975}, especially during 1$^{st}$ transients. However, such differences could be simply driven by the inherent scatter in EP experiments (the results of the 2$^{nd}$ permeation transient CC model, where trapping and surface effects appear to be minimized, suggests a minimum CV of ~8\%) or poor agreement with the actual boundary conditions of the experiment (Fig. \ref{fig:ComparisonMethods}). While leveraging a full fit approach will not solve all these challenges, it will provide insights into whether the use of the CC model-based simple lag time and breakthrough time approaches is justified.

\subsubsection{Sources of error in ITDS experiments}
As shown in Fig. \ref{fig:ComparisonMethods}, hydrogen diffusivity measurements conducted using ITDS have considerably less error than measurements using EP methods. This reduction in error is likely attributed to three factors. First, ITDS uses hydrogen egress from a pre-charged sample to establish the material diffusivity; therefore, the equilibrium hydrogen trap distribution will be reached during the precharging step. This reduces trapping effects during the subsequent desorption experiment and explains the reasonable agreement between the average diffusivities measured with ITDS and the $2^{nd}$ permeation transient. Second, the reduced error in ITDS experiments also manifests from the analysis approach. Specifically, the hydrogen diffusivity is calculated from the gradient in the desorbed hydrogen concentration rate (wppm/s) with time rather than the magnitude of the hydrogen concentration vs time (as happens in EP experiments, where the permeation current is proportional to the hydrogen concentration \cite{Bockris1965,Pound1993,Akiyama2015,Liu2016}). As such, subtle deviations in diffusivity induce noticeable changes in the ITDS fit, resulting in a concomitant increase in sensitivity for this approach.

\begin{figure}[htp]
     \centering
         \centering
         \includegraphics[width=0.85\textwidth]{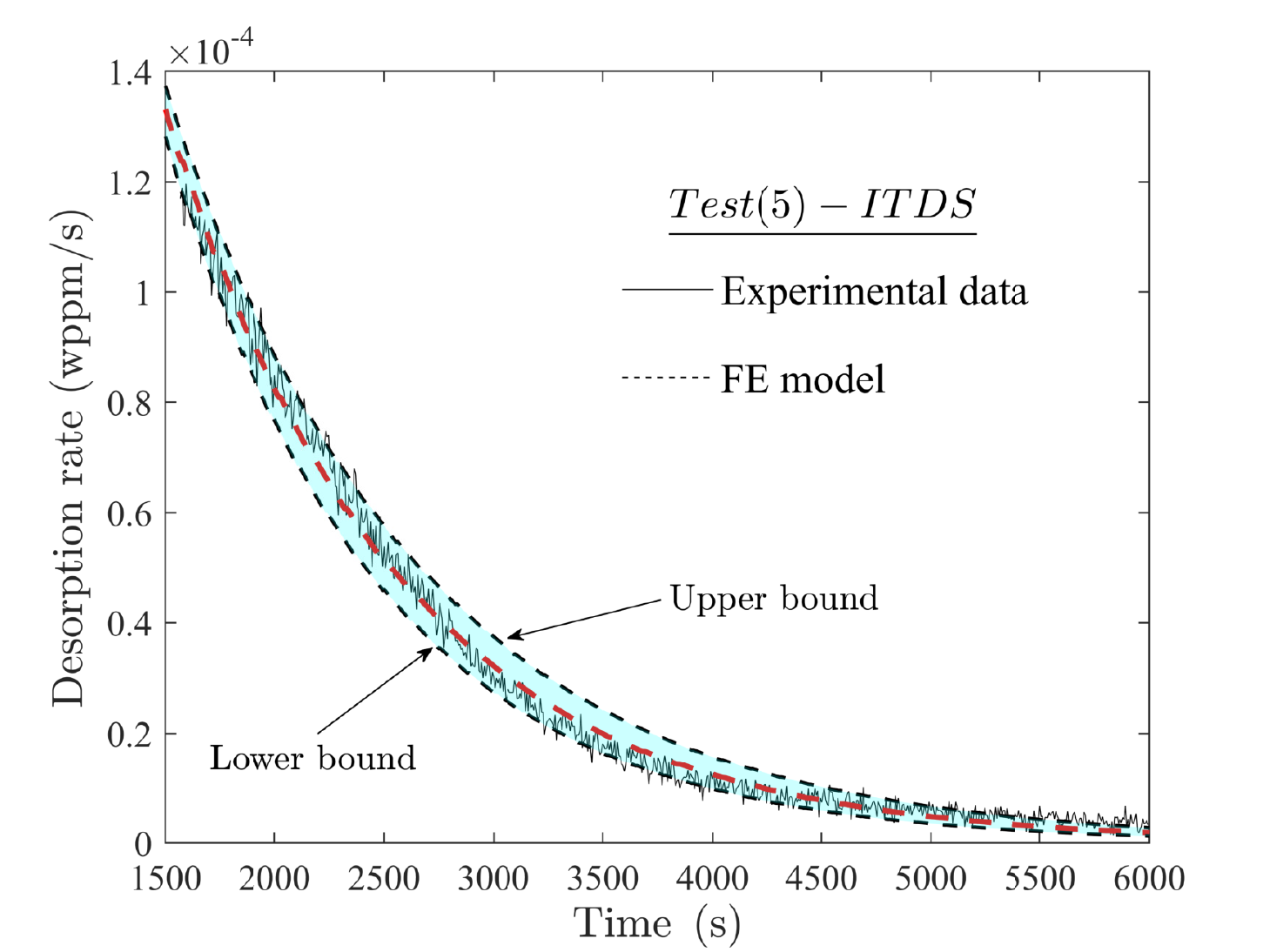}
        \caption{Fitted ITDS curve with lower and upper bounds determined with the lowest coefficient of variation (8.4\%) associated to the second permeation transient (CC model).}
        \label{fig:SensitivityDiscussion}
\end{figure}

To highlight the sensitivity of the ITDS analysis, a representative ITDS desorption curve (Run 5) and the associated COMSOL-Matlab best fit (red dashed line) are plotted in Fig. \ref{fig:SensitivityDiscussion}. The best fit diffusivity (7.72x10$^{-11}$ m$^2$/s) for this experiment was then modified by the CV (8.5\%) from the 2$^{nd}$ EP transient (CC model) to establish  upper (8.36x10$^{-11}$ m$^2$/s) and lower (7.07x10$^{-11}$ m$^2$/s) bound diffusivities. The hydrogen desorption curves for these modified diffusivities were then calculated, keeping $C_{0iso}$ from the original fit (0.71 wppm), to assess how the best-case EP error would affect the ITDS fit results. These upper and lower fits are shown by the black dashed lines in Fig. \ref{fig:SensitivityDiscussion}, which clearly depart from the experimental ITDS data, demonstrating that even an 8\% change in the diffusivity will distinctly affect the goodness of the ITDS fit.

Lastly, the boundary conditions associated with ITDS are well-defined and the majority of the important variables are readily controlled or explicitly known (e.g., specimen thickness, time between the end of charging and the start of the ITDS measurement). In fact, the primary source of scatter in ITDS variable is likely the initial hydrogen concentration ($C_{0iso}$) introduced during the pre-charging. As reported in \cite{Zafra2022}, the measured hydrogen concentration at the end of precharging can vary appreciably, even when significant efforts are made to keep all charging parameters identical. As such, it is important to evaluate whether or not the $C_{0iso}$ value calculated from the ITDS experiment fits is a realistic value. One pathway to assess the reasonableness of $C_{0iso}$ is to compare it to the calculated surface hydrogen concentration ($C_{0app}$) on the entry side of the EP experiment. Since both the ITDS and $1^{st}$ transients used galvanostatic charging at 5 mA/cm$^2$, these two approaches should yield similar values. Regarding $C_{0app}$, it can be calculated as follows:

\begin{equation}\label{eq:C0app}
    C_{0app}= \frac{J_{ss1}L}{D F}
\end{equation}

\noindent where $L$ is the membrane thickness, $D$ the diffusion coefficient, $J_{ss1}$ is the steady state permeation current, and F is Faraday's constant; note that Eq. (\ref{eq:C0app}) assumes the CC model boundary conditions. According to Table \ref{Isothermal}, $C_{0iso}$ is 0.46$\pm$0.22 wppm, while the $C_{0app}$ calculated from Table \ref{PermeationFirst} using the CC method results is 0.59$\pm$0.13 wppm. While the average EP-calculated surface concentration is slightly higher than that calculated for the ITDS, significant overlap in error bars exists between the two approaches, suggesting that the fitted $C_{0iso}$ and therefore the ITDS fitting methodology are reasonable.

\subsection{Implications of results}
The preceding results and discussion provide an opportunity to comment on the general efficacy of the ITDS and EP approaches for determining hydrogen diffusivity. First, the present data confirm the utility of the ITDS approach for assessing hydrogen diffusivity relative to EP methods, consistent with prior work by the authors \cite{Zafra2022}. Despite using a fast diffusing material (cold-rolled pure Fe) in a thin sheet geometry (\textless 1 mm thick), which has historically been incompatible with ITDS \cite{Mine2009}, the ITDS and EP (especially the 2$^{nd}$ transient) experiments yielded similar average hydrogen diffusivities. Additionally, good agreement was observed between the averages of the calculated surface hydrogen concentration in the EP experiments ($C_{0app}$) and the calculated initial hydrogen concentration for the ITDS experiments ($C_{0iso}$). However, while these average metrics were often in reasonable agreement, the scatter in diffusivity from the EP experiments was always at least 2 times higher than that observed in ITDS experiments (Fig. \ref{fig:ComparisonScattering} and Fig. \ref{fig:ComparisonMethods}), irrespective of analysis strategy and permeation transient.

Second, the current results suggest that the increased scatter in EP experiments is likely driven by variations in electrochemical boundary conditions due to surface conditions, with a secondary contribution from hydrogen trapping. Comparison of three different diffusion models for EP revealed a substantial influence (up to five-fold differences) of these boundary conditions on both the average diffusivity (Table \ref{PermeationFitBC}) and scatter (Fig. \ref{fig:ComparisonMethods}). Such results have several practical implications. For example, it is common in the literature to analyze EP data, even results generated with galvanostatic charging, using the conventional lag and breakthrough time methods. However, these analysis strategies are closed form solutions to the CC model, while galvanostatic charging provides a boundary condition more in line with the CF model. This can result in misleading evaluations since the fits for the CC and CF models to EP data can be reasonably similar (Fig. \ref{fig:PermeFits}), but the CF model consistently yields a three-fold higher diffusivity than the CC model. Additionally, if galvanostatic boundary conditions are employed, it should be recognized that the applied electrochemical potential will likely vary from specimen to specimen as well as throughout the experiment due to surface influences, particularly during a 1$^{st}$ transient experiment. These influences will eventually be attenuated \cite{Turnbull1995}, but inherently introduce error into the analysis that will especially influence short permeation experiments (such as the case for fast-diffusing materials like pure Fe). Based on these observations, it is suggested that EP experiments be conducted using potentiostatic approaches. This suggestion is driven by: (1) the fact that the standard lag time and breakthrough time solutions are derived from the CC model, which is best aligned with a potentostatic experiment, (2) the established relationship between applied electrochemical potential and hydrogen fugacity \cite{Subramanyan1981}, which sets the surface hydrogen concentration, and (3) a reduction in surface effect variations since the evolution of the applied potential that occurrs during galvanostatic charging can be avoided.

Lastly, while ITDS appears to be ideal for determining the hydrogen diffusivity, there are limitations to this approach that should be noted. For example, ITDS may be less effective than EP for materials with strong passive films (Al, Ti, etc.) that could hinder hydrogen desorption until sufficiently high temperatures are reached that reduce the oxide. In this instance, electrochemical techniques may be advantageous due to the strong control of all the variables, $i.e.$, they allow one to apply the current density necessary to reduce these oxides. Additionally, ITDS requires the use of specialized ultra-high vacuum capabilities that can be quite expensive. As such, it should be underscored that EP remains an important and useful tool for assessing hydrogen-metal interactions, but it requires special care and attention to detail to ensure that artifacts are not unknowingly introduced.

\section{Conclusions}
\label{Sec:ConcludingRemarks}

A comparison of the hydrogen diffusivity obtained via EP and ITDS methods has been performed leveraging the results of $>$10 replicate experiments for each approach conducted on cold-rolled pure Fe. Based on these results, the following conclusions can be made:

\begin{itemize}
    \item The average diffusivity obtained from a 1$^{st}$ permeation transient using the standard breakthrough and lag time methods was distinctly lower ($\sim$5.8x10$^{-11}$ cm$^2$/s) than that observed using these same methods on a consecutively run 2$^{nd}$ permeation transient ($\sim$8.4x10$^{-11}$ cm$^2$/s). The observed variation across the replicate permeation experiments decreased from 13.6\% to 8.6\% between the 1$^{st}$ and 2$^{nd}$ permeation transient, respectively. Interestingly, the average diffusivity measured from the ITDS experiments was nominally similar to the 2$^{nd}$ permeation transient ($\sim$7.8x10$^{-11}$ cm$^2$/s), but the variation was reduced by nearly half.
    
    \item The efficacy of the ITDS analysis approach is supported by the close agreement between the best-fit initial hydrogen concentration for the ITDS specimen and the hydrogen concentration calculated for the entry surface of the permeation specimen.
    
    \item The increased scatter in the EP measurements relative to the ITDS experiments was attributed to both an influence of hydrogen trapping and the assumed electrochemical boundary conditions in common EP analysis approaches. The similarity in average diffusivity, but strong difference in scatter, between the ITDS and 2$^{nd}$ permeation transient results suggests that trapping likely provides a secondary contribution relative to electrochemical boundary condition effects.
    
    \item The influence of electrochemical boundary conditions on the hydrogen diffusivity calculated from EP experiments was assessed by fitting three different diffusion models to the large set of generated EP data, $i.e.$, constant concentration (CC) constant (CF) and flux continuity (FC). Not only was the average diffusivity sensitive to the applied boundary conditions (up to four-fold difference for the 2$^{nd}$ permeation transient), but the observed scatter also strongly depends on the employed model (CV ranged from 8\% to ~30\% for the 2$^{nd}$ permeation transient). 
    
    \item These data confirm the improved repeatability of ITDS for determining the hydrogen diffusivity in materials amenable to this technique. For materials where EP may be preferred (e.g., those with resilient passive films), the presented results underscore the need for closely controlling the electrochemical boundary conditions.
 
\end{itemize}

Potential avenues for future work include extending the demonstration of the ITDS approach to other materials, including oxide-bearing case studies.

\section{Acknowledgements}
\label{Sec:Acknowledgeoffunding}

The authors acknowledge financial support from the EPSRC (grants EP/V04902X/1, EP/R010161/1 and EP/V009680/1). E. Mart\'{\i}nez-Pa\~neda was additionally supported by an UKRI Future Leaders Fellowship [grant MR/V024124/1].


\small
\bibliographystyle{elsarticle-num}

\end{document}